\documentclass[prd,aps,showpacs,prev,preprintnumbers,floatfix,nofootinbib,twocolumn]{revtex4-1}

\pdfoutput=1

\usepackage{mathrsfs,amsmath,amsthm,amssymb,color,epstopdf,verbatim,enumerate,graphicx}
\usepackage{float}
\usepackage{hyperref}
\hypersetup{colorlinks,linkcolor=cyan,citecolor=cyan,urlcolor =[rgb]{0.0,0.0,0.5}}
\usepackage[nameinlink,noabbrev]{cleveref}
\bibpunct{\color{cyan}[}{\color{cyan}]}{,}{n}{}{;}

\usepackage{ulem}
\usepackage{xcolor}

\begin{document}

\preprint{P.~He \& B.-Q.~Ma, \href{https://doi.org/10.1103/PhysRevD.108.063006}
{Phys.Rev.D 108 (2023) 063006}}

\title{Comprehensive analysis on photon-electron Lorentz-violation parameter plane}
\author{Ping He$\,{}^{a}$}
\author{Bo-Qiang Ma$\,{}^{a,b,c}$}\email{mabq@pku.edu.cn}
\thanks{corresponding author.}
\affiliation{${}^a$School of Physics, Peking University, Beijing 100871, China\\
		${}^b$Center for High Energy Physics, Peking University, Beijing 100871, China\\
		${}^c$Collaborative Innovation Center of Quantum Matter, Beijing, China}
		

\begin{abstract}
Large High Altitude Air Shower Observatory~(LHAASO) opens the window of ultra-high-energy~(UHE) photon detection, broadens the path of testing basic physical concepts such as Lorentz symmetry, and brings the possibility of potential high-energy physical phenomenon research such as photon decay and electron decay. 
Currently, the UHE photons from LHAASO observation set strict constraints on photon and electron Lorentz symmetry violation~(LV) effects. 
To obtain a global impression of the photon-electron LV parameter plane, we make a detailed analysis for photon decay and electron decay.
Our discussion gives the corresponding decay thresholds and energy-momentum distributions in different LV parameter configurations. 
We get corresponding constraints on the photon LV parameter, electron LV parameter and the photon-electron LV parameter plane from LHAASO observation.
For the space allowed for LV effect, that is beyond relativity, we also provide corresponding boundaries from LHAASO observation.
\end{abstract}

\maketitle

\section{Introduction}\label{Introduction}

Large High Altitude Air Shower Observatory~(LHAASO) is a new generation gamma-ray and cosmic-ray observatory with work bands at TeV and PeV energies~\cite{H40-Cao-2010-future,H41-Cao-2019-introduction,H42-Cao-2022-large}. 
LHAASO consists of three detector arrays; a large fraction of detectors started the operation in 2019, and the whole detector construction was finished in 2021.
The commissioning of LHAASO brings ultra-high-energy~(UHE) photon detection into a new stage. 
LHAASO reported more than 530 UHE photons with energies larger than $100~\mathrm{TeV}$ from twelve astrophysical gamma-ray sources within the Milky Way, including the highest-energy photon detected at about $1.42~\mathrm{PeV}$ from the gamma-ray source LHAASO J2032+4102 at the direction of the Cygnus~\cite{H1-cao-2021-ultrahigh}.
LHAASO reported the detection of the Crab Nebula gamma-ray spectrum from $5\times10^{-4}~\mathrm{PeV}$ to $1.12~\mathrm{PeV}$, and these UHE photons 
are shown to exhibit the Crab Nebula as a PeV electron accelerator~\cite{H8-cao-2021-pata}. 
LHAASO discoveries not only help to study the origin and acceleration mechanism of UHE cosmic rays, but also provide the opportunity to test fundamental physics concepts such as Lorentz symmetry~\cite{H3-lhaaso-2021-exploring,H2-li-2021-ultrahigh,H5-chen-2021-strong,H7-li-2022-testing,H43-He-2022-joint}.\\

High-energy particles from the Universe offer the opportunity to detect the highest-energy particles that people can observe. 
LHAASO opens the window of UHE photon detection, broadens the path of testing basic physical concepts such as Lorentz symmetry, and brings the research opportunity for potentially high-energy phenomenon such as photon decay and electron decay. 
In the classic case, photon decay $\gamma\to e^\mathrm{+}+e^\mathrm{-}$ and electron decay $e^\mathrm{\pm}\to e^\mathrm{\pm}+\gamma$ are forbidden, but in the Lorentz symmetry violation~(LV) case, things might be different. 
Photon decay is a photon LV phenomenon proposed by Coleman and Glashow~\cite{H46-Coleman-1997-cosmic, H47-Coleman-1998-high}, and there are strict constraints on the photon superluminal LV scale from the observation of high-energy photons.
If the LV effects make the velocity of electrons bigger than the velocity of photons, electron decay may occur.
It can be analogous to a charged particle propagating in a medium: when the speed of the charged particle is bigger than the light speed in the medium, the charged particle can radiate an electromagnetic field, and this phenomenon is called Cherenkov radiation. 
The electron decay caused by LV can be called electron vacuum Cherenkov radiation.
Early in 2001, Stecker and Glashow used the observation of Mrk 501 to constrain the maximum electron velocity~\cite{H39-Stecker-2001-new}.
Between 2002 and 2006, Jacobson et al.~\cite{H13-Jacobson-2001-TeV,H12-jacobson-2003-threshold,H45-Jacobson-2003-new,H44-Jacobson-2005-Lorentz} studied potential electromagnetic LV phenomena, and offered constraints for photon decay and electron decay.
Jacobson, Liberati and Mattingly~\cite{H51-Jacobson-2002-strong} argued that synchrotron radiation from the Crab Nebula imposes a stringent constraint on any modification of the
dispersion relations of the electron that might be induced by quantum gravity. Ellis, Mavromatos and Sakharov~\cite{H48-Ellis-2003-synchrotron} pointed out further that the photon dispersion relation might have a linear form of LV correction while the dispersion relation of the electron is severely constrained; therefore synchrotron radiation from the Crab Nebula can be used to discriminate between models of space-time foam.
There have also been studies on the LV constraints using data related to Crab Nebula observations (such as Refs.~\cite{H49-MAGIC-2017-constraining,H50-Satunin-2019-new}).
The observation of photon or electron decay would indicate a signature for new physics such as the LV effect to challenge the current physical framework. 
On the other hand, the highest-energy photon and electron set very strict constraints on LV effects.\\

LHAASO observations promote the LV effect constraint to a new stage.
For photon decay, the LHAASO Collaboration got a photon superluminal linear LV constraint $E_\mathrm{LV}^\mathrm{(\gamma,sup)}\ge1.42\times10^{24}~\mathrm{GeV}$ through the analysis of two gamma-ray sources LHAASO J2032+4102 and J0534+2202~\cite{H3-lhaaso-2021-exploring}.
From photon decay analysis, Ref.~\cite{H2-li-2021-ultrahigh} got a photon superluminal linear LV constraint $E_\mathrm{LV}^\mathrm{(\gamma,sup)}\ge2.74\times10^{24}~\mathrm{GeV}$ from the LHAASO $1.42~\mathrm{PeV}$ highest-energy photon, and this constraint, which has an improvement of 2 to 4 orders of magnitude over previous constraints to photon decay, is the strictest constraint that can be gotten from LHAASO observation. 
From photon decay and photon splitting $\gamma\to N\gamma$ research, Ref.~\cite{H5-chen-2021-strong} proposed corresponding constraints for linear and quadratic photon energy scales though the analysis of LHAASO's three sources and an UHE photon event.
As a gamma-ray detector, LHAASO cannot directly obtain the energy of electrons in the astrophysical source, but through detecting the spectrum of the Crab Nebula with detailed analysis, LHAASO shows that the Crab Nebula operates as an electron PeVatron~\cite{H8-cao-2021-pata}.
According to the synchrotron self-Compton model~\cite{H9-atoyan-1996-mechanisms,H10-Meyer-2010-crab},\footnote{The synchrotron self-Compton model is a likely mechanism to produce the observed $1.12~\mathrm{PeV}$ photon, and there are also possibilities that the photon might be produced in some other mechanisms: for example, hadronic interaction of high-energy protons~(or nuclei) via neutral pion decay.} the Crab Nebula high-energy gamma ray is produced via inverse-Compton process by UHE electrons. 
Through systematic analysis, LHAASO got a simple relation between the upscattered photon $E^\mathrm{\gamma}$ and the parent electron $E^\mathrm{e}$: $E^\mathrm{e}=2.15(E^\mathrm{\gamma}/1~\mathrm{PeV})^{0.77}~\mathrm{PeV}$~\cite{H8-cao-2021-pata}. 
Thus, for the Crab Nebula $1.12~\mathrm{PeV}$ highest-energy photon, the energy of the parent electron is $2.3~\mathrm{PeV}$~\cite{H8-cao-2021-pata}, which is the highest-energy electron that we got from LHAASO observation. 
By analyzing the Crab Nebula $1.12~\mathrm{PeV}$ highest-energy photon, Ref.~\cite{H7-li-2022-testing} obtained the most strict constraint on electron superluminal linear scale $ E_\mathrm{LV}^\mathrm{(e,sup)}\ge9.4\times10^{25}~\mathrm{GeV}$, and this constraint improves previous bounds on electron LV effect by $10^4$ times. 
Considering both photon and electron LV effects in photon decay and electron decay research, Ref.~\cite{H43-He-2022-joint} got a joint constraint on the photon-electron LV parameter plane from LHAASO's highest-energy photon and electron, and this joint constraint can naturally derive the strictest constraint on photon and electron LV effects respectively. \\

These are some assumptions in these constraints from LHAASO data. 
For photon decay, usually only the photon LV effect is considered, but the LV effect of the outgoing electron-positron pair is neglected. 
In usual electron decay analysis~\cite{H7-li-2022-testing}, there is a supposition that the emitted photon is soft enough that its LV effect can be neglected. 
Since the outgoing particles obtain the whole energy-momentum of the initial photon, it is necessary to consider the LV effect with more careful consideration. 
Therefore we make a more detailed analysis for photon decay and electron decay by various cases, and discuss the physical implications of these cases. 
We show the corresponding decay thresholds in different parameter configurations and the different energy-momentum distributions in different decay cases. 
We demonstrate how the highest-energy photon and electron constrain the photon LV parameter, electron LV parameter and the photon-electron LV parameter plane.  \\

\section{Research and Discussion}\label{Research and Discussion}

We generally expect that LV effects have observable effects in extremely high energy, and the LV modifications would be suppressed by the Planck scale $E_\text{Pl}\simeq1.22\times10^{19}~\mathrm{GeV}$. 
So we use the LV parameter $\xi_\mathrm{n}$ and $\eta_\mathrm{n}$ to modify the dispersion relation of photons and electrons~(or positrons):
\begin{equation}\label{photon/electron dispresion relation by xi/eta}
    \begin{cases}
    w^2=k^2[1+\xi_\mathrm{n}(\frac{k}{E_\mathrm{Pl}})^n]   &  \mathrm{photon};\\
    E^2=m^2+p^2[1+\eta_\mathrm{n}(\frac{p}{E_\mathrm{Pl}})^n]  &  \mathrm{electron/positron}, \\
    \end{cases}
\end{equation}
where $E_\mathrm{Pl}$ is the Plank scale, and $\xi_\mathrm{n}$, $\eta_\mathrm{n}$ are the $n$th-order LV parameters of photons and electrons respectively.
By convenience, if we only consider the linear modification, we set $\xi_\mathrm{1}\equiv\xi$, $\eta_\mathrm{1}\equiv\eta$. 
For a decay process of a high-energy particle, the threshold occurs when the final particle momenta are parallel to the initial particle momentum~\cite{H6-Mattingly-2002-threshold}. 
So it is reasonable for us to only consider the modulus of the momentum $k=|\vec{k}|$, and $p=|\vec{p}|$.
Choosing this model-independent method, we can study LV phenomena without causing ambiguities from different theory models~\cite{H18-He-2022-lorentz}. \\

Besides photon LV parameter $\xi_\mathrm{n}$ and electron LV parameter $\eta_\mathrm{n}$, it is also common to use the LV scale $E_\mathrm{LV,n}^{\gamma/\mathrm{e}}$ to modify the dispersion relation of photons and electrons~(or positrons):
\begin{equation}\label{photon/electron dispresion relation by E_LV}
    \begin{cases}
    \omega^2=k^2[1-s_\mathrm{n}^\gamma(\frac{k}{E_\mathrm{LV,n}^{\gamma}})^n]   &  \mathrm{photon};\\
    E^2=m^2+p^2[1-s_\mathrm{n}^\mathrm{e}(\frac{p}{E_\mathrm{LV,n}^\mathrm{e}})^n]  &  \mathrm{electron/positron}, \\
    \end{cases}
\end{equation}
where $E_\mathrm{LV,n}^{\gamma/\mathrm{e}}$ are  the hypothetical energy scale of photons/electrons at which the $n$th-order LV effect becomes significant, and the LV scales $E_\mathrm{LV,n}^{\gamma/\mathrm{e}}$ for subluminal and superluminal cases are different. 
$s_\mathrm{n}^{\gamma/\mathrm{e}}=+1$ means that the higher the energy the slower the photon speed~(subluminal), and $s_\mathrm{n}^{\gamma/\mathrm{e}}=-1$ means faster photon speed~(superluminal).
For linear modification, we set $E_\mathrm{LV,1}^{\gamma/\mathrm{e}}\equiv E_\mathrm{LV}^{\gamma/\mathrm{e}}$ and $s_\mathrm{1}^{\gamma/\mathrm{e}}\equiv s^{\gamma/\mathrm{e}}$, and the correspondence between LV parameters and LV scales is
\begin{equation}\label{correspondence between two parameterization}
    \begin{cases}
    \xi\frac{1}{E_\mathrm{Pl}}=-s^{\gamma}\frac{1}{E_\mathrm{LV}^{\gamma}}   &  \mathrm{photon};\\
    \eta\frac{1}{E_\mathrm{Pl}}=-s^\mathrm{e}\frac{1}{E_\mathrm{LV}^\mathrm{e}}  &  \mathrm{electron/positron}. \\
    \end{cases}
\end{equation}

As we have introduced, current astronomical observations allow an opportunity to revisit photon decay and electron decay.
The LHAASO $1.42~\mathrm{PeV}$ highest-energy photon, that is directly observed, and the LHAASO $2.3~\mathrm{PeV}$ highest-energy electron, that is indirectly inferred, set very strict constraints on photon and electron parameters.
Observing high-energy photons/electrons~(whose energy is $E_{\gamma/\mathrm{e}}$) means that $E_{\gamma/\mathrm{e}}$ does not reach the photon/electron decay threshold $k_\mathrm{th}^{\gamma/\mathrm{e}}$: $k_\mathrm{th}^{\gamma/\mathrm{e}}>E_{\gamma/\mathrm{e}}$. 
Next, we discuss the photon decay and electron decay by various cases, and discuss the different physics implications of different cases. \\

\subsection{Photon decay}

LHAASO reported the highest-energy photon at about $1.42~\mathrm{PeV}$ from the gamma-ray source LHAASO J2032+4102 at the direction of the Cygnus~\cite{H1-cao-2021-ultrahigh}. This LHAASO $1.42~\mathrm{PeV}$ highest-energy photon constrains the photon-electron plane by constraining photon decay.
In the classic case, the photon decay $\gamma\to e^\mathrm{+}+e^\mathrm{-}$ is forbidden, but in the LV case, this reaction might occur in some LV parameter configurations. 
If a high-energy photon with momentum $k$ decays into an electron with momentum $xk$~($x\in[0,1]$) and a positron with momentum $(1-x)k$, using the energy-momentum conservation relation, we get
\begin{equation}\label{energy-momentum conservation relation for photon decay I}
 E_\mathrm{photon}(k)=E_\mathrm{electron}(xk)+E_\mathrm{positron}[(1-x)k]. 
\end{equation}
With photon/electron dispersion relation Eq.~(\ref{photon/electron dispresion relation by xi/eta}) expanding to the first order of the LV parameters and to the leading order of $(m/k)^2$, Eq.~(\ref{energy-momentum conservation relation for photon decay I})  reduces to~\cite{H12-jacobson-2003-threshold}
\begin{equation}\label{energy-momentum conservation relation for photon decay II}
k[1+\frac{\xi_\mathrm{n}}{2}(\frac{k}{E_\mathrm{Pl}})^n]=xk[1+\frac{m^2}{2(xk)^2}+\frac{\eta_\mathrm{n}}{2}(\frac{xk}{E_\mathrm{Pl}})^n]+\{x\leftrightarrow 1-x\}.
\end{equation}
After simple algebraic operations, the above formula becomes~\cite{H12-jacobson-2003-threshold}
\begin{equation}\label{energy-momentum conservation relation for photon decay III}
\frac{m^2E^n_\mathrm{Pl}}{k^{n+2}}=x(1-x)[\xi_\mathrm{n}-((1-x)^{n+1}+x^{n+1})\eta_\mathrm{n}].
\end{equation}
Eq.~(\ref{energy-momentum conservation relation for photon decay III}) means that finding the photon decay threshold is equivalent to finding the minimum value of $k$ on the left side of Eq.~(\ref{energy-momentum conservation relation for photon decay III}), and correspondingly, maximizing the right side of Eq.~(\ref{energy-momentum conservation relation for photon decay III}). 
The $x$ value satisfying Eq.~(\ref{energy-momentum conservation relation for photon decay III}) reflects the momentum distribution after photon decay, where $x=1$ (or $0$) means that the electron~(or positron) obtains the most momentum and $x=1/2$ means that the momentum distribution is equal among the electron and the positron. \\

If we only consider the linear~($n=1$) modification, Eq.~(\ref{energy-momentum conservation relation for photon decay III}) becomes~\cite{H12-jacobson-2003-threshold}
\begin{equation}\label{e-m relation of photon decay for linear modification}
    \frac{m^2E_\mathrm{Pl}}{k^3}=x(1-x)[\xi-((1-x)^2+x^2)\eta].
\end{equation}
When $\xi\to0$ and $\eta\to0$, the maximum value of the right side tends to zero, and $k_\mathrm{th}\to +\infty$.
This corresponds to the situation where photons cannot decay in the classic case as expected, since any situation must return back to the classical case when the LV effects approach zero.
Next let us see how the highest-energy photon $E_{\gamma}$ sets constraints on LV linear modification parameters under different cases.

\begin{itemize}

    \item Case I. $\xi\ne0, \eta=0$\\
    Case I means that there is only photon linear LV modification; then Eq.~(\ref{e-m relation of photon decay for linear modification}) becomes~\cite{H11-shao-2010-lorentz}
    \begin{equation}\label{e-m relation of photon decay for linear modification, eta=0}
     \frac{m^2E_\mathrm{Pl}}{k^3}=x(1-x)\xi.   
    \end{equation}
    \begin{enumerate}
    
        \item	 
        When $\xi<0$, the right side of Eq.~(\ref{e-m relation of photon decay for linear modification, eta=0}) cannot be positive within the allowed range of values, so there is no photon decay, i.e., the photon decay threshold is
        \begin{equation}\label{threshold of photon decay, eta=0, xi<0}
             k_\mathrm{th}^{\gamma}=+\infty.
         \end{equation}
         In this situation, it is always $k_\mathrm{th}^{\gamma}>E_{\gamma}$, so the highest-energy photon does not set extra constraints.
         
         \item	
         When $\xi>0$, the maximum value on the right side of Eq.~(\ref{e-m relation of photon decay for linear modification, eta=0}) is $\xi/4$, and the corresponding photon decay threshold is~\cite{H12-jacobson-2003-threshold,H11-shao-2010-lorentz}
         \begin{equation}\label{threshold of photon decay, eta=0, xi>0}
             k_\mathrm{th}^{\gamma}=(\frac{4m^2E_\mathrm{Pl}}{\xi})^{1/3}.
         \end{equation}
         This threshold is taken at $x=1/2$, i.e., the momenta of the outgoing particles are equally distributed. 
         If photon decay does occur in this situation, there is a superluminal linear LV modification for photons. 
         Conversely, finding a high-energy photon~(whose energy reaches $E_{\gamma}$) means that $E_{\gamma}$ dose not reach the threshold $k_\mathrm{th}^{\gamma}>E_{\gamma}$. 
         The highest-energy photon sets a constraint on photon LV parameter $0<\xi<4m^2E_\mathrm{Pl}/E_{\gamma}^3$, which corresponds to a superluminal constraint ($s^{\gamma}=-1$) for the photon LV scale $E_\mathrm{LV}^{\gamma,\mathrm{sup}}>E_{\gamma}^3/(4m^2)$. 
         Considering the LHAASO $1.42~\mathrm{PeV}$ highest-energy photon, we get $0<\xi<4.45\times10^{-6}$ and $E_\mathrm{LV}^{\gamma,\mathrm{sup}}>2.74\times10^{24}~\mathrm{GeV}$~\footnote{Considering the experimental errors of the LHAASO highest-energy photon $1.42^{+0.13}_{-0.13}~\mathrm{PeV}$, we get $0<\xi<4.45^{-1.03}_{+1.49}\times10^{-6}$ and $E_\mathrm{LV}^{\gamma,\mathrm{sup}}>2.74^{+0.83}_{-0.68}\times10^{24}~\mathrm{GeV}$.}, which is the same as the strictest photon superluminal LV constraint~\cite{H2-li-2021-ultrahigh}. 
         This same result is natural as the strictest constraint is gotten under the same assumption that $\eta=0$.
    \end{enumerate}

    \item Case II. $\xi=0, \eta\ne0$\\
    Case II corresponds to the situation with only electron linear LV modification; then Eq.~(\ref{e-m relation of photon decay for linear modification}) becomes
    \begin{equation}\label{e-m relation of photon decay for linear modification, xi=0}
     \frac{m^2E_\mathrm{Pl}}{k^3}=x(x-1)[(1-x)^2+x^2]\eta. 
    \end{equation}
    \begin{enumerate}
    
        \item	
        When $\eta>0$, the right side of Eq.~(\ref{e-m relation of photon decay for linear modification, xi=0}) is always less than zero within the allowed range of values, i.e., there is no photon decay. We can take the photon decay threshold as
        \begin{equation}\label{threshold of photon decay, eta>0, xi=0}
             k_\mathrm{th}^{\gamma}=+\infty.
         \end{equation}
         In this situation, it is always $k_\mathrm{th}^{\gamma}>E_{\gamma}$, so the highest-energy photon does not set extra constraints.
         
         \item	
         When $\eta<0$, the maximum value on the right side of Eq.~(\ref{e-m relation of photon decay for linear modification, xi=0}) is $-\eta/8$, and the photon decay threshold is
         \begin{equation}\label{threshold of photon decay, eta<0, xi=0}
             k_\mathrm{th}^{\gamma}=(\frac{-8m^2E_\mathrm{Pl}}{\eta})^{1/3}.
         \end{equation}
         This threshold is taken at $x=1/2$, i.e., the momenta of the outgoing electron and positron are equally distributed. 
         If photon decay does occur in this situation, there is a subluminal linear LV modification for electrons. 
         Conversely, finding the highest-energy photon sets a constraint on electron LV parameter $-8m^2E_\mathrm{Pl}/E_{\gamma}^3<\eta<0$, which corresponds to a subluminal constraint~($s^\mathrm{e}=+1$) for electron LV scale $E_\mathrm{LV}^\mathrm{e, sub}>E_{\gamma}^3/(8m^2)$. 
         Considering the LHAASO $1.42~\mathrm{PeV}$ highest-energy photon, we get $-8.90\times10^{-6}<\eta<0$ and $E_\mathrm{LV}^\mathrm{e, sub}>1.37\times10^{24}~\mathrm{GeV}$~\footnote{Considering the experimental errors of the LHAASO highest-energy photon $1.42^{+0.13}_{-0.13}~\mathrm{PeV}$, we get $-8.90^{+2.06}_{-2.97}\times10^{-6}<\eta<0$ and $E_\mathrm{LV}^\mathrm{e, sub}>1.37^{+0.41}_{-0.34}\times10^{24}~\mathrm{GeV}$.}.
    \end{enumerate}

    \item Case III. $\xi=\eta\ne0$\\
    Case III corresponds to the situation that the photon and electron LV modification parameters are the same and both are linear modifications; then Eq.~(\ref{e-m relation of photon decay for linear modification}) becomes~\cite{H11-shao-2010-lorentz}
    \begin{equation}\label{e-m relation of photon decay for linear modification, xi=eta/=0}
     \frac{m^2E_\mathrm{Pl}}{k^3}=x(1-x)[1-(1-x)^2-x^2]\xi.   
    \end{equation}
    \begin{enumerate}
    
        \item	
        When $\xi=\eta<0$, the right side of Eq.~(\ref{e-m relation of photon decay for linear modification, xi=eta/=0}) is always negative~($\le0$) within the range of values, so there is no photon decay, i.e., the photon decay threshold is
        \begin{equation}\label{threshold of photon decay, eta=xi<0}
             k_\mathrm{th}^{\gamma}=+\infty.
         \end{equation}
         In this situation, it is always $k_\mathrm{th}^{\gamma}>E_{\gamma}$, so the highest-energy photon does not set extra constraints.
         
         \item	
         When $\xi=\eta>0$, the maximum value on the right side of Eq.~(\ref{e-m relation of photon decay for linear modification, xi=eta/=0}) is $\xi/8$, and the photon decay threshold is~\cite{H12-jacobson-2003-threshold,H11-shao-2010-lorentz}
         \begin{equation}\label{threshold of photon decay, eta=xi>0}
             k_\mathrm{th}^{\gamma}=(\frac{8m^2E_\mathrm{Pl}}{\xi})^{1/3}.
         \end{equation}
         This threshold is taken at $x=1/2$, i.e., the momenta of the outgoing particles are equally distributed. 
         If photon decay does occur in this situation, there are superluminal linear modifications for both photons and electrons. 
         Conversely, finding the highest-energy photon sets constraints on photon and electron LV parameter $0<\xi=\eta<8m^2E_\mathrm{Pl}/E_{\gamma}^3$, which corresponds to superluminal constraints~($s^{\gamma}=s^\mathrm{e}=-1$) for both the photon and electron LV scale $E_\mathrm{LV}^{\gamma,\mathrm{sup}}=E_\mathrm{LV}^\mathrm{e, sup}>E_{\gamma}^3/(8m^2)$.
         Considering the LHAASO $1.42~\mathrm{PeV}$ highest-energy photon, we get $0<\xi=\eta<8.90\times10^{-6}$ and $E_\mathrm{LV}^{\gamma,\mathrm{sup}}=E_\mathrm{LV}^\mathrm{e, sup}>1.37\times10^{24}~\mathrm{GeV}$~\footnote{Considering the experimental errors of the LHAASO highest-energy photon $1.42^{+0.13}_{-0.13}~\mathrm{PeV}$, we get $0<\xi=\eta<8.90^{-2.06}_{+2.97}\times10^{-6}$ and $E_\mathrm{LV}^{\gamma,\mathrm{sup}}=E_\mathrm{LV}^\mathrm{e, sup}>1.37^{+0.41}_{-0.34}\times10^{24}~\mathrm{GeV}$.}.
    \end{enumerate}

    \item Case IV. $\xi\ne0, \eta\ne0$\\
    Case IV corresponds to the most general situation: the electron and photon LV parameters take arbitrary values. Introducing a new variable $z=(2x-1)^2$ can make the analysis simple, so that $x=(1+\sqrt{z})/2$, $(1-x)=(1-\sqrt{z})/2$ and $x(1-x)=(1-z)/4$~\cite{H12-jacobson-2003-threshold}. 
    The relevant range of $z$ is $[0,1]$, where $z=0$ corresponds to the symmetric configuration $x=1/2$ and $z=1$ corresponds to $x=1~\mathrm{or}~0$~\cite{H12-jacobson-2003-threshold}. 
    In terms of $z$, Eq.~(\ref{e-m relation of photon decay for linear modification}) becomes~\cite{H12-jacobson-2003-threshold}
     \begin{equation}\label{e-m relation of photon decay for linear modification, eta/=0, xi/=0, z}
     \frac{m^2E_\mathrm{Pl}}{k^3}=\frac{1-z}{4}\xi-\frac{1-z^2}{8}\eta.
    \end{equation}
    \begin{enumerate}
    
        \item	
        When $\xi-\eta<0$ and $2\xi-\eta<0$, the right side of Eq.~(\ref{e-m relation of photon decay for linear modification, eta/=0, xi/=0, z}) is less than zero~(i.e., $\le0$) within the allowed range of values, so there is no photon decay~\cite{H12-jacobson-2003-threshold,H13-Jacobson-2001-TeV}, i.e., the photon decay threshold is
        \begin{equation}\label{threshold of photon decay, eta>xi, eta-2xi>0}
             k_\mathrm{th}^{\gamma}=+\infty.
         \end{equation}
         In this situation, it is always $k_\mathrm{th}^{\gamma}>E_{\gamma}$, so the highest-energy photon does not set extra constraints.
         
         \item	
         When $\xi>0$ and $2\xi-\eta>0$, the maximum value on the right side of Eq.~(\ref{e-m relation of photon decay for linear modification, eta/=0, xi/=0, z}) is $(2\xi-\eta)/8$, and the photon decay threshold is~\cite{H12-jacobson-2003-threshold,H13-Jacobson-2001-TeV}
         \begin{equation}\label{threshold of photon decay, xi>0, 2xi-eta>0}
             k_\mathrm{th}^{\gamma}=(\frac{8m^2E_\mathrm{Pl}}{2\xi-\eta})^{1/3}.
         \end{equation}
         This threshold is taken at $z=0$, that is $x=1/2$~\cite{H12-jacobson-2003-threshold} when the momenta of the outgoing particles generated by photon decay are equally distributed.
         In this situation, $k_\mathrm{th}^{\gamma}>E_{\gamma}$ means $0<2\xi-\eta<8m^2E_\mathrm{Pl}/E_{\gamma}^3$~\cite{H12-jacobson-2003-threshold}. 
         Considering the LHAASO $1.42~\mathrm{PeV}$ highest-energy photon, we get $0<2\xi-\eta<8.90\times10^{-6}$.
    
          \item	
          When $\eta<\xi<0$, the maximum value on the right side of Eq.~(\ref{e-m relation of photon decay for linear modification, eta/=0, xi/=0, z}) is $-(\xi-\eta)^2/(8\eta)$, and the photon decay threshold is~\cite{H12-jacobson-2003-threshold,H13-Jacobson-2001-TeV}
         \begin{equation}\label{threshold of photon decay, eta<xi<0}
             k_\mathrm{th}^{\gamma}=(\frac{-8\eta m^2E_\mathrm{Pl}}{(\xi-\eta)^2})^{1/3}.
         \end{equation}
         This threshold is taken at $z=\xi/\eta$~\cite{H12-jacobson-2003-threshold}, that is $x=(1+\sqrt{\xi/\eta})/2$.
         In this situation, $k_\mathrm{th}^{\gamma}>E_{\gamma}$ means $0<\xi-\eta<\sqrt{-8m^2E_\mathrm{Pl}\eta/E_{\gamma}^3}$~\cite{H12-jacobson-2003-threshold}. 
         Considering the LHAASO $1.42~\mathrm{PeV}$ highest-energy photon, we get $0<\xi-\eta<\sqrt{-8.90\times10^{-6}\eta}$.
     \end{enumerate}
         
     \begin{figure}[H]
    \centering
    \includegraphics[scale=0.5]{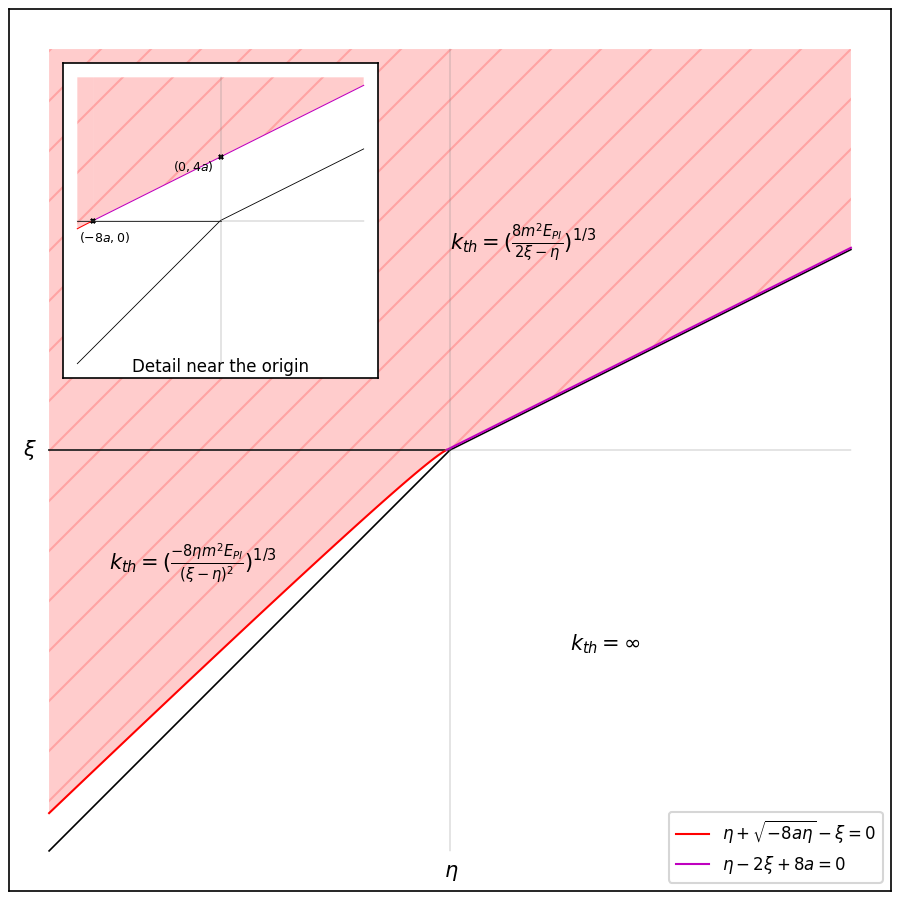}
    \caption{Photon decay constraint on the photon-electron LV parameter plane from highest-energy photon~($a:\equiv m^2E_\mathrm{Pl}/E_{\gamma}^3)$). In global figure the value range of horizontal and vertical coordinates is $-10^{-3}\sim10^{-3}$, and in the detail figure around the origin the value range of horizontal and vertical coordinates is $-10^{-5}\sim10^{-5}$.}
    \label{fig:photon decay}
     \end{figure}

\end{itemize}

Fig.~1 shows the constraints from the high-energy photon, where the shaded portion is the prohibited parameter space.
The discussion of photon decay $\gamma\to e^++e^-$ can be extended to other outgoing charged particles $\gamma\to \mathrm{X}^++\mathrm{X}^-$~\cite{H29-Altschul-2006-astrophysical}. 
The difference is that the electron mass should be changed to another outgoing particle mass. Since the photon energy must exceed the outgoing particle energy, in usual discussion we just consider the outgoing particles as an electron-positron pair. \\

\subsection{Electron decay}

The Crab Nebula gamma-ray spectrum detected by LHAASO ranges from $5\times10^{-4}~\mathrm{PeV}$ to $1.12~\mathrm{PeV}$, and these UHE photons demonstrate that the Crab Nebula operates as an electron PeVatron~\cite{H8-cao-2021-pata}. 
As we introduced above, by adopting the synchrotron self-Compton model~\cite{H9-atoyan-1996-mechanisms,H10-Meyer-2010-crab}, LHAASO determined that the $1.12~\mathrm{PeV}$ upper limit energy of the Crab Nebula gamma-ray spectrum means that the energy of the parent electron in the Crab Nebula reaches $2.3~\mathrm{PeV}$~\cite{H8-cao-2021-pata}.
The synchrotron self-Compton model implies that the Crab Nebula gamma ray above $1~\mathrm{GeV}$ is produced via inverse-Compton process by UHE electrons. 
In the analysis of the inverse-Compton process, there is no LV effect being considered, as the inverse-Compton process is hardly affected by LV effect due to a very tiny variation in the allowed phase space for the cross section~\cite{H7-li-2022-testing}. 
Next, we use the Crab Nebula $2.3~\mathrm{PeV}$ electron indirectly obtained from LHAASO to constrain the photon-electron LV parameter plane by constraining electron decay.
In the classic case, the electron decay $e^\mathrm{-}\to e^\mathrm{-}+\gamma$ is forbidden, but in the LV case, this reaction might occur in some LV parameter configurations. 
When a high-energy electron with momentum $k$ decays into an electron with momentum $yk$~($y\in[0,1]$) and a photon with momentum $(1-y)k$, using photon/electron dispersion relation Eq.~(\ref{photon/electron dispresion relation by xi/eta}) and the energy-momentum conservation relation, we get
\begin{equation}\label{energy-momentum conservation relation for electron decay I}
 E_\mathrm{electron}(k)=E_\mathrm{electron}(yk)+E_\mathrm{photon}[(1-y)k].  
\end{equation}
With expansion to the first order of the LV parameters and to the leading order of $(m/k)^2$, Eq.~(\ref{energy-momentum conservation relation for electron decay I}) reduces to
\begin{equation}\label{energy-momentum conservation relation for electron decay II}
   \begin{aligned}
    k[1+\frac{m^2}{2k^2}+\frac{\eta_\mathrm{n}}{2}(\frac{k}{E_\mathrm{Pl}})^n]=yk[1+\frac{m^2}{2(yk)^2}+\frac{\eta_\mathrm{n}}{2}(\frac{yk}{E_\mathrm{Pl}})^n]\\+(1-y)k[1+\frac{\xi_\mathrm{n}}{2}(\frac{(1-y)k}{E_\mathrm{Pl}})^n].
   \end{aligned}
\end{equation}
After simple algebraic operations, the above formula becomes~\cite{H12-jacobson-2003-threshold}
\begin{equation}\label{energy-momentum conservation relation for electron decay III}
    \frac{m^2E_\mathrm{Pl}^n}{k^{n+2}}=\frac{y^{n+2}-y}{y-1}\eta_\mathrm{n}-y(1-y)^n\xi_\mathrm{n},
\end{equation}
which means that finding the electron decay threshold is equivalent to finding the minimum value of $k$ on the left side of Eq.~(\ref{energy-momentum conservation relation for electron decay III}), and correspondingly, maximizing the right side of Eq.~(\ref{energy-momentum conservation relation for electron decay III}). 
The $y$ value satisfying Eq.~(\ref{energy-momentum conservation relation for electron decay III}) reflects the momentum distribution after electron decay, where $y=1$ (or $0$) means that the electron~(or photon) obtains the most momentum and $y=1/2$ means that the momentum distribution is equal among the electron and the photon after the electron decay.\\

If we only consider the linear~($n=1$) modification, Eq.~(\ref{energy-momentum conservation relation for electron decay III}) becomes~\cite{H12-jacobson-2003-threshold}
\begin{equation}\label{e-m relation of electron decay for linear modification}
    \frac{m^2E_\mathrm{Pl}}{k^3}=y(y+1)\eta-y(1-y)\xi.
\end{equation}
When $\xi\to0$ and $\eta\to0$, the maximum value of the right side tends to zero, and $k_\mathrm{th}\to +\infty$. 
This corresponds to the situation where electrons cannot decay in the classic case.  
Next let us see how the highest-energy electron $E_\mathrm{e}$ sets constraints on LV linear modification parameters in different cases.

\begin{itemize}

    \item Case I. $\xi\ne0, \eta=0$\\
    Case I means that there is only photon linear LV modification; then Eq.~(\ref{e-m relation of electron decay for linear modification}) becomes
    \begin{equation}\label{e-m relation of electron decay for linear modification, eta=0}
     \frac{m^2E_\mathrm{Pl}}{k^3}=y(y-1)\xi.   
    \end{equation}
    \begin{enumerate}
    
        \item
        When $\xi>0$, the right side of Eq.~(\ref{e-m relation of electron decay for linear modification, eta=0}) is less than zero~($\le0$) within the allowed range of values, so there is no electron decay, i.e., the electron decay threshold is
        \begin{equation}\label{threshold of electron decay, eta=0,xi>0}
             k_\mathrm{th}^\mathrm{e}=+\infty.
         \end{equation}
         In this situation, it is always $k_\mathrm{th}^\mathrm{e}>E_\mathrm{e}$, so the highest-energy electron does not set extra constraints.
         
         \item	
         When $\xi<0$, the maximum value on the right side of Eq.~(\ref{e-m relation of electron decay for linear modification, eta=0}) is $-\xi/4$, and the electron decay threshold is
         \begin{equation}\label{threshold of electron decay, eta=0, xi<0}
             k_\mathrm{th}^\mathrm{e}=(\frac{-4m^2E_\mathrm{Pl}}{\xi})^{1/3}.
         \end{equation}
         This threshold is taken at $y=1/2$.
         If electron decay does occur in this situation, there is a subluminal linear LV modification for the photon. 
         Conversely, finding a high-energy electron~(whose energy reaches $E_\mathrm{e}$) means that $E_\mathrm{e}$ does not reach the threshold $k_\mathrm{th}^\mathrm{e}>E_\mathrm{e}$. 
         The highest-energy electron sets a constraint on the photon LV parameter $-4m^2E_\mathrm{Pl}/E_\mathrm{e}^3<\xi<0$, which corresponds to a subluminal constraint~($s^{\gamma}=+1$) for the photon LV scale $E_\mathrm{LV}^{\gamma,\mathrm{sub}}>E_\mathrm{e}^3/(4m^2)$.
         Considering the LHAASO $2.3~\mathrm{PeV}$ highest-energy electron, we get $-1.0\times10^{-6}<\xi<0$ and $E_\mathrm{LV}^{\gamma,\mathrm{sub}}>1.2\times10^{25}~\mathrm{GeV}$.\\
    \end{enumerate}

    \item Case II. $\xi=0, \eta\ne0$\\
    Case II means that there is only electron linear LV modification; then Eq.~(\ref{e-m relation of electron decay for linear modification}) becomes
    \begin{equation}\label{e-m relation of electron decay for linear modification, xi=0}
     \frac{m^2E_\mathrm{Pl}}{k^3}=y(y+1)\eta.   
    \end{equation}
    \begin{enumerate}
    
        \item	
        When $\eta<0$, the right side of Eq.~(\ref{e-m relation of electron decay for linear modification, xi=0}) is negative ($\le0$) within the allowed range of values, so there is no electron decay. The electron decay threshold is
        \begin{equation}\label{threshold of electron decay, eta<0, xi=0}
             k_\mathrm{th}^\mathrm{e}=+\infty.
         \end{equation}
         In this situation, it is always $k_\mathrm{th}^\mathrm{e}>E_\mathrm{e}$, so the highest-energy electron does not set extra constraints.
         
         \item	
         When $\eta>0$, the maximum value on the right side of Eq.~(\ref{e-m relation of electron decay for linear modification, xi=0}) is $2\eta$, and the electron decay threshold is
         \begin{equation}\label{threshold of electron decay, eta>0, xi=0}
             k_\mathrm{th}^\mathrm{e}=(\frac{m^2E_\mathrm{Pl}}{2\eta})^{1/3}.
         \end{equation}
         This threshold is taken at $y=1$, i.e., the outgoing electron gains almost all of the momentum of the parent electron.
         If electron decay does occur in this situation, there is a superluminal linear LV modification for the electron. 
         Conversely, finding the highest-energy electron sets a constraint on electron LV parameter $0<\eta<m^2E_\mathrm{Pl}/(2E_\mathrm{e}^3)$, which corresponds to a superluminal constraint~($s^\mathrm{e}=-1$) for the electron LV scale $E_\mathrm{LV}^\mathrm{e, sup}>2E_\mathrm{e}^3/m^2$. 
         Considering the LHAASO $2.3~\mathrm{PeV}$ highest-energy electron, we get $0<\eta<1.3\times10^{-7}$ and $E_\mathrm{LV}^\mathrm{e, sup}>9.4\times10^{25}~\mathrm{GeV}$, which is the same as the strictest electron superluminal LV constraint~\cite{H7-li-2022-testing}.
         This same result is natural as the strictest constraint is gotten under the same assumption of $\xi=0$.
    \end{enumerate}

    \item Case III. $\xi=\eta\ne0$\\
    Case III corresponds to a simple assumption: the photon and electron LV modification parameters are the same, and both are linear modifications, and then Eq.~(\ref{e-m relation of electron decay for linear modification}) becomes
    \begin{equation}\label{e-m relation of electron decay for linear modification, eta=xi/=0}
     \frac{m^2E_\mathrm{Pl}}{k^3}=2y^2\xi.   
    \end{equation}
    \begin{enumerate}
    
        \item	
        When $\xi=\eta<0$, the right side of Eq.~(\ref{e-m relation of electron decay for linear modification, eta=xi/=0}) is no more than zero~($\le0$) within the range of values, and there is no electron decay, that is, the electron decay threshold is
        \begin{equation}\label{threshold of electron decay, eta=xi<0}
             k_\mathrm{th}^\mathrm{e}=+\infty.
         \end{equation}
         In this situation, it is always $k_\mathrm{th}^\mathrm{e}>E_\mathrm{e}$, so the highest-energy electron does not set extra constraints.
         
         \item	
         When $\xi=\eta>0$, the maximum value on the right side of Eq.~(\ref{e-m relation of electron decay for linear modification, eta=xi/=0}) is $2\xi$, and the electron decay threshold is
         \begin{equation}\label{threshold of electron decay, eta=xi>0}
            k_\mathrm{th}^\mathrm{e}=(\frac{m^2E_\mathrm{Pl}}{2\xi})^{1/3}.
         \end{equation}
         This threshold is taken at $y=1$, that is, the outgoing electron gains almost all of the momentum of the parent electron.
         If electron decay does occur in this situation, there are superluminal linear modifications for both the photon and electron. 
         Conversely, finding the highest-energy electron sets constraints on both the photon and electron LV parameter $0<\xi=\eta<m^2E_\mathrm{Pl}/(2E_\mathrm{e}^3)$, and they are superluminal constraints~($s^{\gamma}=s^\mathrm{e}=-1$) for both the photon and electron LV scale $E_\mathrm{LV}^{\gamma,\mathrm{sup}}=E_\mathrm{LV}^\mathrm{e, sup}>2E_\mathrm{e}^3/m^2$. 
         Considering the LHAASO $2.3~\mathrm{PeV}$ highest-energy electron, we get $0<\xi=\eta<1.3\times10^{-7}$ and $E_\mathrm{LV}^{\gamma,\mathrm{sup}}=E_\mathrm{LV}^\mathrm{e, sup}>9.4\times10^{25}~\mathrm{GeV}$.
    \end{enumerate}

    \item Case IV. $\xi\ne0, \eta\ne0$\\
    Case IV corresponds to the most general situation: the photon and electron LV parameters take arbitrary values. After simple algebraic operations, we get the following:
    \begin{enumerate}
    
        \item	
        When $\eta<0$ and $\xi-\eta>0$, the right side of Eq.~(\ref{e-m relation of electron decay for linear modification}) is negative ($\le0$) within the range of values. 
        Thus there is no electron decay~\cite{H12-jacobson-2003-threshold}, i.e., the electron decay threshold is
        \begin{equation}\label{threshold of electron decay, eta<0,eta-xi<0}
             k_\mathrm{th}^\mathrm{e}=+\infty.
         \end{equation}
         In this situation, it is always $k_\mathrm{th}^\mathrm{e}>E_\mathrm{e}$, so the highest-energy electron does not set extra constraints.
         
         \item	
         When $\eta>0$ and $\xi+3\eta>0$, the maximum value on the right side of Eq.~(\ref{e-m relation of electron decay for linear modification}) is $2\eta$, and the electron decay threshold is~\cite{H12-jacobson-2003-threshold,H13-Jacobson-2001-TeV}
         \begin{equation}\label{threshold of electron decay, eta>0, 3eta+xi>0}
             k_\mathrm{th}^\mathrm{e}=(\frac{m^2E_\mathrm{Pl}}{2\eta})^{1/3}.
         \end{equation}
         This threshold is taken at $y=1$.
         In this situation, $k_\mathrm{th}^\mathrm{e}>E_\mathrm{e}$ means $0<\eta<m^2E_\mathrm{Pl}/(2E_\mathrm{e}^3)$~\cite{H12-jacobson-2003-threshold}. 
         Considering the LHAASO $2.3~\mathrm{PeV}$ highest-energy electron, we get $0<\eta<1.3\times10^{-7}$.
         
          \item	
          When $\xi-\eta<0$ and $\xi+3\eta<0$, the maximum value on the right side of Eq.~(\ref{e-m relation of electron decay for linear modification}) is $-(\xi-\eta)^2/(4\xi+4\eta)$, and the electron decay threshold is~\cite{H12-jacobson-2003-threshold,H13-Jacobson-2001-TeV}
         \begin{equation}\label{threshold of electron decay, eta-xi>0, 3eta+xi>0}
             k_\mathrm{th}^\mathrm{e}=(\frac{-4(\xi+\eta)m^2E_\mathrm{Pl}}{(\xi-\eta)^2})^{1/3}.
         \end{equation}
         This threshold is taken at $y=(\xi-\eta)/(2\xi+2\eta)$~\cite{H12-jacobson-2003-threshold}. 
         In this situation, $k_\mathrm{th}^\mathrm{e}>E_\mathrm{e}$ means $-\sqrt{-4m^2E_\mathrm{Pl}(\eta+\xi)/E_\mathrm{e}^3}<\xi-\eta<0$~\cite{H12-jacobson-2003-threshold}. 
         Considering the LHAASO $2.3~\mathrm{PeV}$ highest-energy electron, we get $-\sqrt{-1.0\times10^{-6}(\eta+\xi)}<\xi-\eta<0$.
    \end{enumerate}

    \begin{figure}[H]
    \centering
    \includegraphics[scale=0.5]{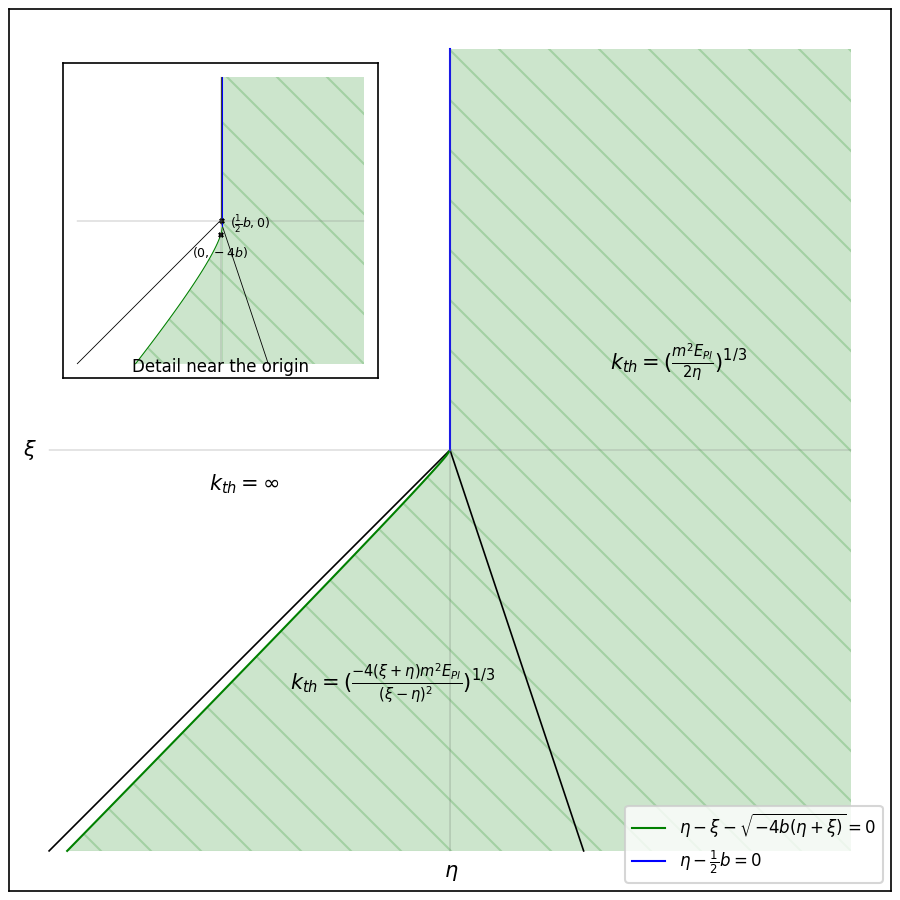}
    \caption{Electron decay constraint on photon-electron LV parameter plane from highest-energy electron~($b:\equiv m^2E_\mathrm{Pl}/E_\mathrm{e}^3)$). In the global figure the value range of horizontal and vertical coordinates is $-10^{-3}\sim10^{-3}$, and in the detail figure around the origin the value range of horizontal and vertical coordinates is $-10^{-5}\sim10^{-5}$.}
    \label{fig:electron decay}
     \end{figure}

\end{itemize}

The constraints from the high-energy electron are shown in Fig.~2, where the shaded portion is the prohibited parameter space.

\subsection{Joint constraint from electron decay and photon decay}

\begin{figure}[H]
    \centering
    \includegraphics[scale=0.5]{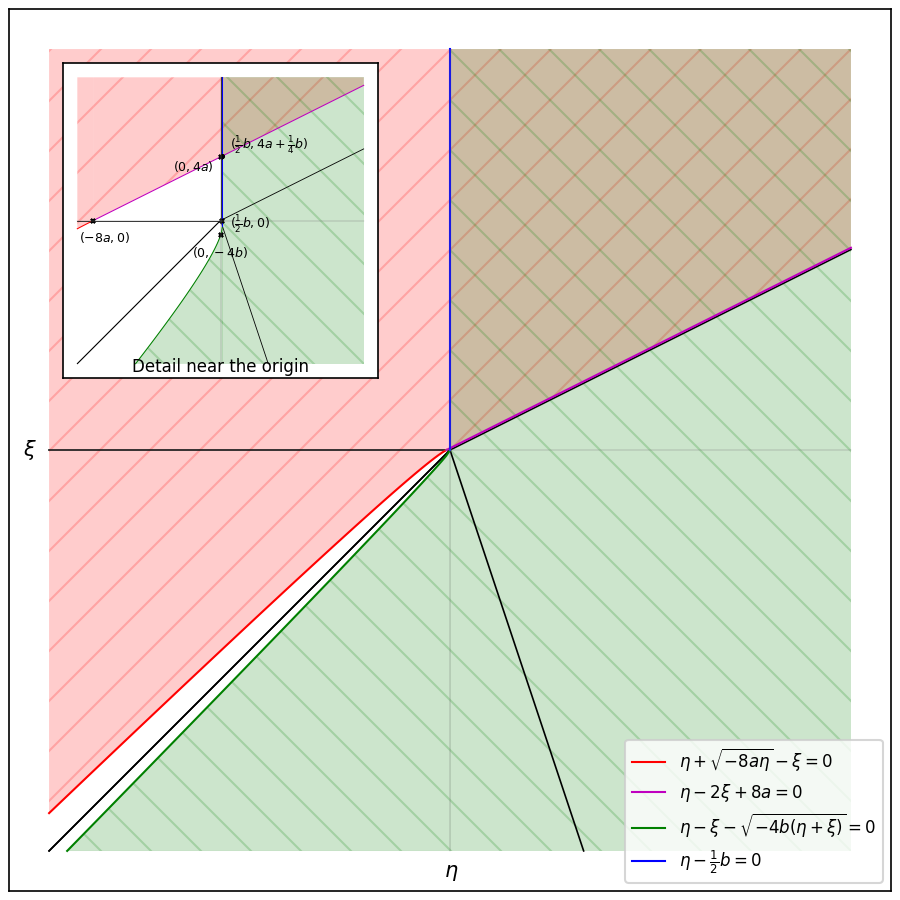}
    \caption{Joint constraint on photon-electron LV parameter plane from highest-energy photon and electron~($a:\equiv m^2E_\mathrm{Pl}/E_{\gamma}^3)$ and $b:\equiv m^2E_\mathrm{Pl}/E_\mathrm{e}^3)$). In the global figure the value range of horizontal and vertical coordinates is $-10^{-3}\sim10^{-3}$, and in the detail figure around the origin the value range of horizontal and vertical coordinates is $-10^{-5}\sim10^{-5}$.}
    \label{fig:joint limitation}
\end{figure}

In the above discussion, we have discussed how highest-energy photons and electrons constrain the photon-electron LV parameter plane by electron decay and photon decay respectively. In the follow discussion, we show the joint constraint on the photon-electron LV parameter plane. To see more clearly, we show the joint constraint in Fig.~\ref{fig:joint limitation}.

If we presuppose that there is no electron LV effect ($\eta=0$), the constraint on the photon LV parameter is $-4m^2E_\mathrm{Pl}/E_\mathrm{e}^3<\xi<4m^2E_\mathrm{Pl}/E_{\gamma}^3$.
The lower limitation $-4m^2E_\mathrm{Pl}/E_\mathrm{e}^3<\xi$ is from the highest-energy electron, and it is a subluminal constraint~($s^{\gamma}=+1$) for the photon LV scale $E_\mathrm{LV}^{\gamma,\mathrm{sub}}>E_\mathrm{e}^3/(4m^2)$. 
The upper limitation $\xi<4m^2E_\mathrm{Pl}/E_{\gamma}^3$ is from the highest-energy photon, and it is a superluminal constraint~($s^{\gamma}=-1$) for the photon LV scale $E_\mathrm{LV}^{\gamma,\mathrm{sup}}>E_{\gamma}^3/(4m^2)$. 
Considering the LHAASO $1.42~\mathrm{PeV}$ highest-energy photon and $2.3~\mathrm{PeV}$ highest-energy electron, we get $-1.0\times10^{-6}<\xi<4.45\times10^{-6}$, which corresponds to $E_\mathrm{LV}^{\gamma,\mathrm{sub}}>1.2\times10^{25}~\mathrm{GeV}$ and $E_\mathrm{LV}^{\gamma,\mathrm{sup}}>2.74\times10^{24}~\mathrm{GeV}$.
We note that $E_\mathrm{LV}^{\gamma,\mathrm{sup}}>2.74\times10^{24}~\mathrm{GeV}$ is the same as the strictest photon superluminal constraint from the highest-energy photon~\cite{H2-li-2021-ultrahigh}, as this strictest constraint is got under the the same assumption of $\eta=0$.\\

If we presuppose that there is no photon LV effect ($\xi=0$), the constraint on the electron LV parameter is $-8m^2E_\mathrm{Pl}/E_{\gamma}^3<\eta<m^2E_\mathrm{Pl}/(2E_\mathrm{e}^3)$.
The lower limitation $-8m^2E_\mathrm{Pl}/E_{\gamma}^3<\eta$ is from the highest-energy photon, and it is a subluminal constraint~($s^\mathrm{e}=+1$) for the electron LV scale $E_\mathrm{LV}^\mathrm{e, sub}>E_{\gamma}^3/(8m^2)$. 
The upper limitation $\eta<m^2E_\mathrm{Pl}/(2E_\mathrm{e}^3)$ is from the highest-energy electron, and it is a superluminal constraint~($s^\mathrm{e}=-1$) for the electron LV scale $E_\mathrm{LV}^\mathrm{e, sup}>2E_\mathrm{e}^3/m^2$. 
Considering the LHAASO $1.42~\mathrm{PeV}$ highest-energy photon and $2.3~\mathrm{PeV}$ highest-energy electron, we get $-8.90\times10^{-6}<\eta<1.3\times10^{-7}$, which is equivalent to $E_\mathrm{LV}^\mathrm{e, sub}>1.37\times10^{24}~\mathrm{GeV}$ and $E_\mathrm{LV}^\mathrm{e, sup}>9.4\times10^{25}~\mathrm{GeV}$. 
We note that the $E_\mathrm{LV}^\mathrm{e, sup}>9.4\times10^{25}~\mathrm{GeV}$ is the same as the strictest electron superluminal constraint from the highest-energy electron~\cite{H7-li-2022-testing}, as this strictest constraint is gotten under the same assumption of $\xi=0$.\\

The highest-energy photon and electron set a very strict constraint on the first, second and fourth quadrants of the photon-electron plane. 
The limitation pole is $(\eta, \xi)<(m^2E_\mathrm{Pl}/(2E_\mathrm{e}^3), m^2E_\mathrm{Pl}/(4E_\mathrm{e}^3)+4m^2E_\mathrm{Pl}/E_{\gamma}^3)$, which corresponds to the constraint on the superluminal LV scale $(E_\mathrm{LV}^\mathrm{e, sup}, E_\mathrm{LV}^{\gamma,\mathrm{sup}})>(2E_\mathrm{e}^3/m^2, 4E_{\gamma}^3E_\mathrm{e}^3/(16m^2E_\mathrm{e}^3+m^2E_{\gamma}^3))$.
Considering the LHAASO $1.42~\mathrm{PeV}$ highest-energy photon and $2.3~\mathrm{PeV}$ highest-energy electron, we get $(\eta, \xi)<(1.3\times10^{-7}, 4.5\times10^{-6})$ and $(E_\mathrm{LV}^\mathrm{e, sup}, E_\mathrm{LV}^{\gamma,\mathrm{sup}})>(9.4\times10^{25}~\mathrm{GeV}, 2.7\times10^{24}~\mathrm{GeV})$.  \\

In the third quadrant, both the photon and electron are subluminal modifications~($s^{\gamma}=+1$ and $s^\mathrm{e}=+1$), and this parameter space needs more discussion. 
To get the quantitative result, we discuss the third quadrant constraint boundary:
\begin{equation}\label{constraint boundary}
    \begin{cases}
    \eta+\sqrt{-8a\eta}-\xi=0  &  (1*) \mathrm{\quad from \quad photon};\\
    \eta-\sqrt{-4b(\eta+\xi)}-\xi=0  &  (2*) \mathrm{\quad from \quad electron},\\
    \end{cases}
\end{equation}
where $a:\equiv m^2E_\mathrm{Pl}/E_{\gamma}^3$ and $b:\equiv m^2E_\mathrm{Pl}/E_\mathrm{e}^3$. 
From Eq.~(37-1*), we get $\eta-\xi=-\sqrt{-8a\eta}>0$, and from Eq.~(37-2*), we get $\eta-\xi=\sqrt{-4b(\eta+\xi)}<0$. 
So these two boundary curves have no cusp in the third quadrant, and this trend can be seen in Fig.~\ref{fig:joint limitation}. 
From Fig.~\ref{fig:joint limitation}, we get that the alternative parameter space is distributed in very small range, nearby the line $\eta=\xi$. 
This distribution means that the parameters of photons and electrons are of similar magnitudes.\\

If we get a concrete LV parameter of photons~(or electrons), the constraint for the electron~(or photon) parameter changes accordingly. 
For example, if there is a concrete photon parameter $\xi_0$, from the constraint boundary Eq.~(\ref{constraint boundary}), we get the constraint for electrons:
\begin{equation}\label{eta constraint}
    \xi_0-4a-2\sqrt{4a^2-2a\xi_0}<\eta<\xi_0-2b+2\sqrt{b^2-2b\xi_0}.
\end{equation}
If we suppose that the photon LV scale is at $10^{24}~\mathrm{GeV}$, that is $\xi_0\sim-10^{-5}$, considering the LHAASO highest-energy photon and electron, we get the constraint of electrons: $-2.49\times10^{-5}<\eta<-5.93\times10^{-6}$. 
By this example, we know that the LV parameters of photons and electrons are at the similar order, and this result is accordant with the result from Fig.~\ref{fig:joint limitation}. 
For the case that there is a concrete electron parameter $\eta_0$, the constraint of the photon parameter is
\begin{equation}\label{xi constraint}
    \eta_0-2b-2\sqrt{b^2-2b\eta_0}<\xi<\eta_0+2\sqrt{-2a\eta_0}.
\end{equation}
For $\eta_0\sim-10^{-5}$, the constraint for the photon is $-1.51\times10^{-5}<\xi<-0.58\times10^{-6}$.\\

From Fig.~3 and the quantitative analysis above, we know that the allowed space in the third quadrant is concentrated within a small range around $\xi=\eta$. 
It means that the photon velocity and electron maximum velocity are allowed to undergo similar corrections in the subluminal direction.
From the photon/electron dispersion relation Eq.~(\ref{photon/electron dispresion relation by xi/eta}), we know that the correction of the dispersion relation from the LV effect means the correction of photon velocity and electron maximum velocity:
\begin{equation}\label{photon/electron velocity correction}
    \begin{cases}
    v_{\gamma}=\frac{\partial E}{\partial k}\approx c_0[1+\xi\frac{k}{E_\mathrm{Pl}}]   &  \mathrm{photon};\\
    v_\mathrm{e}=\frac{\partial E}{\partial p}\approx c_0[1-\frac{m^2}{2p^2}+\eta\frac{p}{E_\mathrm{Pl}}]  &  \mathrm{electron/positron}, \\
    \end{cases}
\end{equation}
where $c_0$ is the speed of light in the classical case, $v_{\gamma}$ is the photon velocity under the LV effect, and $v_\mathrm{e}$ is the speed of electrons. 
When the electron momentum is very large, we can ignore the electron mass term, and the electron maximum velocity is $v_\mathrm{e}^\mathrm{max}\approx c_0[1+\eta\frac{p}{E_\mathrm{Pl}}]$. 
From Fig.~3, only LV parameters near $\xi=\eta$ in the third quadrant are allowed. It means that if there are obvious LV effects, the photon velocity variation can only be subluminal (decreasing with increasing momentum), and the electron maximum velocity should be also subluminal. 
At the same time, both photon velocity and electron maximum velocity variations are of similar magnitudes.

\section{Conclusion}\label{conclusion}

Restudying the photon decay and electron decay under different cases, we know the corresponding decay thresholds and energy-momentum distributions in different LV parameter configurations. 
From the energy-momentum distributions of the outgoing particles, we know that it is necessary to consider both photon and electron LV effects. 
In different LV parameter configurations, photon decay and electron decay mean different LV modifications, and the LHAASO observations imply different constraints on the photon LV parameter, electron LV parameter and the photon-electron LV parameter plane.
There are strict constraints on the first, second and fourth quadrants of the photon-electron plane, and the LV energy scales are constrained at $10^{24\sim25}~\mathrm{GeV}$, which are $10^{5\sim6}$ times higher than the Planck energy. 
There is still possible parameter space for new physics beyond relativity in the third quadrant, but the LHAASO observation also sets strict boundaries for this space. 
The allowable space for LV parameters is only near $\xi=\eta$ in the third quadrant, which means that the photon and electron LV parameters are only allowed to be subluminal with similar magnitudes. \\

This work is supported by the National Natural Science Foundation of China (Grants No.~12075003 and No.~12335006).


\begin{thebibliography}{99}

\bibitem{H40-Cao-2010-future}
Z.~Cao [LHAASO Collaboration],
``A future project at Tibet: The large high altitude air shower observatory (LHAASO),''
\href{https://doi.org/10.1088/1674-1137/34/2/018}{Chin. Phys. C {\bf 34}, 249-252 (2010)}.


\bibitem{H41-Cao-2019-introduction}
Z.~Cao {\it et al.} [LHAASO Collaboration],
``Introduction to Large High Altitude Air Shower Observatory (LHAASO),''
\href{https://doi.org/10.1016/j.chinastron.2019.11.001}{Chin. Astron. Astrophys. {\bf 43}, 457-478 (2019)}.


\bibitem{H42-Cao-2022-large}
Z.~Cao {\it et al.} [LHAASO Collaboration],
``The large high altitude air shower observatory (LHAASO) science book (2021 Edition),''
{Chin. Phys. C {\bf 46}, 030001-030007 (2022)} [\href{https://arxiv.org/abs/1905.02773}{\tt arXiv:1905.02773}].


\bibitem{H1-cao-2021-ultrahigh}
Z.~Cao {\it et al.} [LHAASO Collaboration],
``Ultrahigh-energy photons up to 1.4 petaelectronvolts from 12 $\gamma$-ray Galactic sources,''
\href{https://doi.org/10.1038/s41586-021-03498-z}{Nature {\bf 594}, no.7861, 33-36 (2021)}.


\bibitem{H8-cao-2021-pata}
Z.~Cao {\it et al.} [LHAASO Collaboration],
``Peta-electron volt gamma-ray emission from the Crab Nebula,''
\href{https://doi.org/10.1126/science.abg5137}{Science {\bf 373}, no.6553, 425-430 (2021)} [\href{https://arxiv.org/abs/2111.06545}{\tt arXiv:2111.06545}].


\bibitem{H3-lhaaso-2021-exploring}
Z.~Cao {\it et al.} [LHAASO Collaboration],
``Exploring Lorentz invariance violation from ultrahigh-energy $\gamma$ rays observed by LHAASO,''
\href{https://doi.org/10.1103/PhysRevLett.128.051102}{Phys. Rev. Lett. {\bf 128}, no.5, 051102 (2022)} [\href{https://arxiv.org/abs/2106.12350}{\tt arXiv:2106.12350}].


\bibitem{H2-li-2021-ultrahigh}
C.~Li and B.-Q.~Ma,
``Ultrahigh-energy photons from LHAASO as probes of Lorentz symmetry violations,''
\href{https://doi.org/10.1103/PhysRevD.104.063012}{Phys. Rev. D {\bf 104}, no.6, 063012 (2021)}
[\href{https://arxiv.org/abs/2105.07967}{\tt arXiv:2105.07967}].

\bibitem{H5-chen-2021-strong}
L.~Chen, Z.~Xiong, C.~Li, S.~Chen and H.~He,
``Strong constraints on Lorentz violation using new $\gamma$-ray observations around PeV,''
\href{https://doi.org/10.1088/1674-1137/ac1166}{Chin. Phys. C {\bf 45}, no.10, 105105 (2021)} [\href{https://arxiv.org/abs/2105.07927}{\tt arXiv:2105.07927}].


\bibitem{H7-li-2022-testing}
C.~Li and B.-Q.~Ma,
``Testing Lorentz invariance of electrons with LHAASO observations of PeV gamma-rays from the Crab Nebula,''
\href{https://doi.org/10.1016/j.physletb.2022.137034}{Phys. Lett. B {\bf 829}, 137034 (2022)} [\href{https://arxiv.org/abs/2204.02956}{\tt arXiv:2204.02956}].


\bibitem{H43-He-2022-joint}
P.~He and B.-Q.~Ma,
``Joint photon-electron Lorentz violation parameter plane from LHAASO data,''
\href{https://doi.org/10.1016/j.physletb.2022.137536}{Phys. Lett. B {\bf 835}, 137536 (2022)} [\href{https://arxiv.org/abs/2210.14817}{\tt arXiv:2210.14817}].


\bibitem{H46-Coleman-1997-cosmic}
S.~R.~Coleman and S.~L.~Glashow,
``Cosmic ray and neutrino tests of special relativity,''
\href{https://doi.org/10.1016/S0370-2693(97)00638-2}{Phys. Lett. B {\bf 405}, 249-252 (1997)} [\href{https://arxiv.org/abs/hep-ph/9703240}{\tt arXiv:hep-ph/9703240}].


\bibitem{H47-Coleman-1998-high}
S.~R.~Coleman and S.~L.~Glashow,
``High-energy tests of Lorentz invariance,''
\href{https://doi.org/10.1103/PhysRevD.59.116008}{Phys. Rev. D {\bf 59}, 116008 (1999)} [\href{https://arxiv.org/abs/hep-ph/9812418}{\tt arXiv:hep-ph/9812418}].


\bibitem{H39-Stecker-2001-new}
F.~W.~Stecker and S.~L.~Glashow,
``New tests of Lorentz invariance following from observations of the highest energy cosmic $\gamma$-rays,''
\href{https://doi.org/10.1016/S0927-6505(01)00137-2}{Astropart. Phys. {\bf 16}, 97-99 (2001)} [\href{https://arxiv.org/abs/astro-ph/0102226}{\tt arXiv:astro-ph/0102226}].


\bibitem{H13-Jacobson-2001-TeV}
T.~Jacobson, S.~Liberati and D.~Mattingly,
``TeV astrophysics constraints on Planck scale Lorentz violation,''
\href{https://doi.org/10.1103/PhysRevD.66.081302}{Phys. Rev. D {\bf 66}, 081302(R) (2002)} [\href{https://arxiv.org/abs/hep-ph/0112207}{\tt arXiv:hep-ph/0112207}].


\bibitem{H12-jacobson-2003-threshold}
T.~Jacobson, S.~Liberati and D.~Mattingly,
``Threshold effects and Planck scale Lorentz violation: combined constraints from high energy astrophysics,''
\href{https://doi.org/10.1103/PhysRevD.67.124011}{Phys. Rev. D {\bf 67}, 124011 (2003)} [\href{https://arxiv.org/abs/hep-ph/0209264}{\tt arXiv:hep-ph/0209264}].


\bibitem{H45-Jacobson-2003-new}
T.~Jacobson, S.~Liberati, D.~Mattingly and F.~W.~Stecker,
``New limits on Planck scale Lorentz violation in QED,''
\href{https://doi.org/10.1103/PhysRevLett.93.021101}{Phys. Rev. Lett. {\bf 93}, 021101 (2004)} [\href{https://arxiv.org/abs/astro-ph/0309681}{\tt arXiv:astro-ph/0309681}].


\bibitem{H44-Jacobson-2005-Lorentz}
T.~Jacobson, S.~Liberati and D.~Mattingly,
``Lorentz violation at high energy: Concepts, phenomena and astrophysical constraints,''
\href{https://doi.org/10.1016/j.aop.2005.06.004}{Annals Phys. {\bf 321}, 150--196 (2006)} [\href{https://arxiv.org/abs/astro-ph/0505267}{\tt arXiv:astro-ph/0505267}].



\bibitem{H51-Jacobson-2002-strong}
T.~Jacobson, S.~Liberati and D.~Mattingly,
``A strong astrophysical constraint on the violation of special relativity by quantum gravity,''
\href{https://doi.org/10.1038/nature01882}{Nature {\bf 424}, 1019-1021 (2003)} 
[\href{https://arxiv.org/abs/astro-ph/0212190}{\tt arXiv:astro-ph/0212190}].


\bibitem{H48-Ellis-2003-synchrotron}
J.~R.~Ellis, N.~E.~Mavromatos and A.~S.~Sakharov,
``Synchrotron radiation from the Crab Nebula discriminates between models of space - time foam,''
\href{https://doi.org/10.1016/j.astropartphys.2003.12.001}{Astropart. Phys. {\bf 20}, 669-682 (2004)} 
[\href{https://arxiv.org/abs/astro-ph/0308403}{\tt arXiv:astro-ph/0308403}].


\bibitem{H49-MAGIC-2017-constraining}
M.~L.~Ahnen \textit{et al.} [MAGIC],
``Constraining Lorentz invariance violation using the Crab Pulsar emission observed up to TeV energies by MAGIC,''
\href{https://doi.org/10.3847/1538-4365/aa8404}{Astrophys. J. Suppl. {\bf 232}, 9 (2017)} 
[\href{https://arxiv.org/abs/1709.00346}{\tt arXiv:1709.00346}].


\bibitem{H50-Satunin-2019-new}
P.~Satunin,
``New constraints on Lorentz Invariance violation from Crab Nebula spectrum beyond $100$ TeV,''
\href{https://doi.org/10.1140/epjc/s10052-019-7520-y}{Eur. Phys. J. C {\bf 79}, 1011 (2019)} 
[\href{https://arxiv.org/abs/1906.08221}{\tt arXiv:1906.08221}].



\bibitem{H9-atoyan-1996-mechanisms}
A.~Atoyan and F.~Aharonian,
``On the mechanisms of gamma radiation in the Crab Nebula,''
{Monthly Notices of the Royal Astronomical Society {\bf 278}, 525-541 (1996)} .


\bibitem{H10-Meyer-2010-crab}
M.~Meyer, D.~Horns and H.~S.~Zechlin,
``The Crab Nebula as a standard candle in very high-energy astrophysics,''
\href{https://doi.org/10.1051/0004-6361/201014108}{Astron. Astrophys. {\bf 523}, A2 (2010)} [\href{https://arxiv.org/abs/1008.4524}{\tt arXiv:1008.4524}].


\bibitem{H6-Mattingly-2002-threshold}
D.~Mattingly, T.~Jacobson and S.~Liberati,
``Threshold configurations in the presence of Lorentz violating dispersion relations,''
\href{https://doi.org/10.1103/PhysRevD.67.124012}{Phys. Rev. D {\bf 67}, 124012 (2003)} [\href{https://arxiv.org/abs/hep-ph/0211466}{\tt arXiv:hep-ph/0211466}].


\bibitem{H18-He-2022-lorentz}
P.~He and B.-Q.~Ma,
``Lorentz Symmetry Violation of Cosmic Photons,''
\href{https://doi.org/10.3390/universe8060323}{Universe {\bf 8}, no.6, 323 (2022)} [\href{https://arxiv.org/abs/2206.08180}{\tt arXiv:2206.08180}].


\bibitem{H11-shao-2010-lorentz}
L.~Shao and B.-Q.~Ma,
``Lorentz violation effects on astrophysical propagation of very high energy photons,''
\href{https://doi.org/10.1142/S0217732310034572}{Mod. Phys. Lett. A {\bf 25}, 3251-3266 (2010)} [\href{https://arxiv.org/abs/1007.2269}{\tt arXiv:1007.2269}].


\bibitem{H29-Altschul-2006-astrophysical}
B.~Altschul,
``Astrophysical limits on Lorentz violation for all charged species,''
\href{https://doi.org/10.1016/j.astropartphys.2007.08.003}{Astropart. Phys. {\bf 28}, 380-384 (2007)} [\href{https://arxiv.org/abs/hep-ph/0610324}{\tt arXiv:hep-ph/0610324}].



\end{thebibliography}
\end{document}